\documentclass[intlimits,twoside,a4paper]{article}

\usepackage[cp1251]{inputenc}

\usepackage[eqsecnum]{cmpj3}
\usepackage{bm}

\usepackage{graphics}
\usepackage{epsfig}
\usepackage{pstricks}
\usepackage{xcolor}
\usepackage{amsmath}
\usepackage{slashed}
\usepackage[normalem]{ulem}
\newcommand \beq{\begin{eqnarray}}
\newcommand \eeq{\end{eqnarray}}
\newcommand \bea{\begin{eqnarray}}
\newcommand \eea{\end{eqnarray}}

\newcommand \kvec{{\bf k}}
\newcommand \qvec{{\bf q}}
\newcommand \pvec{{\bf p}}

\newcommand\rvec{{\bf r}}
\newcommand\tvec{{\bf t}}

\newcommand\Rvec{{\bf R}}

\newcommand\Gvec{{\bf G}}

\definecolor{gold}{rgb}{0.83, 0.69, 0.22}

\issue{2023}{26}{3}{33701}
\doinumber{10.5488/CMP.26.33701}
\title[Neutral band gap of carbon by quantum Monte Carlo
methods]%
{Neutral band gap of carbon by quantum Monte Carlo
methods}%

\author[V. Gorelov, Y. Yang, M. Ruggeri, D. M. Ceperley, C. Pierleoni, M.~Holzmann]
{V. Gorelov\orcid{0000-0003-4324-506X}\refaddr{label1,label2}\thanks{Corresponding author: \email{vitaly.gorelov@polytechnique.edu}.},
 Y. Yang\orcid{0000-0002-8800-9426} \refaddr{label3},
 M. Ruggeri \refaddr{label4},
 D. M. Ceperley \refaddr{label5},
 C. Pierleoni\orcid{0000-0001-9188-3846}\refaddr{label6},
 M.~Holzmann \refaddr{label7}
 }
\addresses{
\addr{label1} LSI, CNRS, CEA/DRF/IRAMIS, \'Ecole Polytechnique, Institut Polytechnique de Paris, F-91120 Palaiseau, France
\addr{label2} European Theoretical Spectroscopy Facility (ETSF)
\addr{label3} Center for Computational Quantum Physics, Flatiron Institute, New York, NY, 10010, USA
\addr{label4} Maison de la Simulation, CEA, CNRS, Univ. Paris-Sud, UVSQ, Universit{\'e} Paris-Saclay, 91191 Gif-sur-Yvette, France
\addr{label5} Department of Physics, University of Illinois Urbana-Champaign, Urbana, Illinois 61801, USA
\addr{label6} Department of Physical and Chemical Sciences, University of L'Aquila, Via Vetoio 10, I-67010 L'Aquila, Italy
\addr{label7} Univ. Grenoble Alpes, CNRS, LPMMC, 38000 Grenoble, France
}


\date{Received March 31, 2023, in final form May 2, 2023}

\begin{document}

\maketitle

\begin{abstract}
We present a method of calculating the energy gap of a charge-neutral excitation using only ground-state calculations. 
We report  Quantum Monte Carlo calculations of $\Gamma\rightarrow\Gamma$ and $\Gamma\rightarrow X$ particle-hole excitation energies in diamond carbon. We analyze the finite-size effect and find the same $1/L$ decay rate as that in a charged excitation, where L is the linear extension of the supercell. This slow decay is attributed to the delocalized nature of the excitation in supercells too small to accommodate excitonic binding effects.
At larger system sizes, the apparent $1/L$ decay crosses over to a $1/L^3$ behavior. Estimation of the
scale of exciton binding can be used to correct finite-size effects of neutral gaps.

\printkeywords
%
\end{abstract}

\section{Introduction}
Quantitative predictions
of electronic excitation spectra in solids has been a central topic
from the very beginning of condensed matter theory. 
A first insight is usually provided by the band structure obtained by density functional
theory (DFT).
Many-body perturbation theory has successfully 
extended DFT or self-consistent field approaches to provide electronic
excitation spectra including perturbative
electronic interaction effects. However, the precise values of spectral 
quantities like the band gap are still sensitive to the underlying 
functional and approximation scheme \cite{book}.
Their  accuracy 
is frequently established by comparison with experiment involving further 
modelling.
Based on the variational principle,
quantum Monte Carlo (QMC) methods \cite{rev1,rev2,rev3,Dubecky2020} provide an alternative way 
to obtain ground and excited state properties 
whose quality  may be judged independently of experimental measurements.

Particularly adapted to study electronic correlation effects in liquids and solids,
QMC calculations for extended systems have become more and more affordable.
Similar to many classical simulation methods, QMC calculations are done
in  a finite (periodic) simulation cell of linear extension $L$, containing $N_e$ electrons. 
Within projector Monte Carlo methods, e.g., fixed-node diffusion Monte Carlo (DMC) \cite{FN},
the systematic bias is entirely determined
by the nodes of the underlying many-body wave function.
Backflow \cite{BF0}, $n-$body and iterated backflow \cite{nbody,BFiter} and related neural network wave functions
\cite{Max,MatthewEG} have
offered a systematic way of improving the accuracy judged by the variational principle.
Results in the thermodynamic limit are then obtained by numerical extrapolation of different system sizes,
possibly accelerated via analytical correction terms.
Dominant sources of finite-size errrors in electronic systems are 
fermionic shell \cite{Lin01} and Coulomb charge effects \cite{fse}.
Since the studies of a fixed-node error are in general limited to small system sizes,
the control of both of these systematic biases,
fixed-node and finite-size error, will eventually
determine the overall accuracy of QMC calculations.

Recently, we have shown that QMC calculations of the fundamental (charged) gap of an $N_e$ electron system, 
frequently called quasi-particle (QP) gap, defined
by the electron addition and removal energies,
\beq
\Delta_{QP} =E_0(N_e+1)+E_0(N_e-1)-2 E_0(N_e),
\label{DeltaQP}
\eeq
 converge slowly, inversely proportional
to the linear extension $L \sim N_e^{-1/3}$ of the simulation cell \cite{bandgap}. 
Heuristically, the $1/L$ dependance results from the interaction of the additional charges
across the periodic boundaries \cite{Payne,Engel} 
which should be absent
in neutral systems. Then, finite-size effects would be drastically reduced
in calculations of the neutral gap,
\beq
\Delta_n = E_1(N_e)-E_0(N_e),
\label{DeltaN}
\eeq
obtained by exciting the system at a fixed number of electrons.
Therefore, QMC calculations of charge neutral, particle hole excitations
\cite{Ceperley87,Mitas94,Williamson98,Towler2000,Kolorenc08,Ma13,Wagner14,Yu15,Zheng18,Frank19,
Neuscamman,Lubos21} 
commonly assume a faster $1/L^3$ behavior to extrapolate finite-size errors.
However,  QMC results of charged and neutral gaps
of diamond Si \cite{Hunt,Lubos21} yields comparable values
at fixed finite system size, not supporting the expected qualitative change of finite-size effects.

In this paper, we study finite-size effects of QMC calculations of the neutral gap.
Similar to~\cite{bandgap}, we explain that leading-order finite-size effects are encoded
in the asymptotic behavior of the static structure factor, $S(\kvec)$, which is encoded in the many-body
wave function underlying the calculation. The size effects are not caused by the charged nature
of the excitation, but by their extended or localized character.
In particular, promoting a single Bloch orbital
from the ground state
to an excited state in the determinantal part of the correlated wave function,
leads to an excitation of extended character, thus it suffers from the same
slow $1/L$ decay as their charged counterparts.

In section~\ref{sec2} we briefly introduce our theoretical methods
which allows us to obtain the neutral gap as a ground-state calculation, discuss a possible use of twisted boundary conditions, and present the general form of the trial wave functions. In section~\ref{sec3} we present our QMC results for diamond carbon compared with quasiparticle-gap from~\cite{bandgap}. In the next section~\ref{sec4} we discuss finite-size effects and we finally recollect our conclusions and perspectives in the final section~\ref{sec5}.

\section{Methodology}\label{sec2}

\subsection{Neutral gap}

The neutral gap as defined above, equation~(\ref{DeltaN}), 
explicitly involves the energy of the first excited state. In practice,
QMC methods of excited state are either based on exact symmetry constraints \cite{Raja95,Foulkes99},
or are approximately implemented in a minimization procedure \cite{Lucas,Neuscamman}.
However, any contamination of the excited state by the ground state will violate the
variational principle, and may lead to a bias for the gap. Here, we show how to restate the
neutral gap within a pure ground-state formalism.

For ideal solid structures, the Hamiltonian is
invariant under the simultaneous translation
of all electrons by any crystal lattice vector, $\tvec$. Therefore,
electronic energies, $E_{\kvec}$, and eigenstates, $\Psi_{\kvec}$,
can be labelled by a wave vector 
$\kvec$ in the first Brillouin zone in the reciprocal crystal lattice.
Wave functions of a given crystal momentum $\kvec$ should satisfy
\beq
\Psi_{\kvec}(\rvec_1+ \tvec,\rvec_2+ \tvec,\dots,\rvec_{N_e}+\tvec)
= \re^{\ri \kvec \cdot \tvec}  \Psi_{\kvec}(\rvec_1,\rvec_2,\dots,\rvec_{N_e}).
\label{Psik}
\eeq
In general, we can assume that the ground state is in the trivial representation, $\kvec=0$,
with energy $E_0(N_e)$.
In an insulator, all excitations of the ground state
are gapped,
$E_{\kvec}(N_e) - E_{0}(N_e) \geqslant \Delta_n$ for $\kvec \ne 0$.
Approaching the thermodynamic limit, $\kvec$ becomes a continuous parameter
and we can write
\beq
\Delta_n = \min_{\kvec \ne 0} E_{\kvec}(N_e) - E_{0}(N_e).
\label{DeltaNkvec}
\eeq
Therefore, similar to the fundamental gap \cite{bandgap}, also the neutral gap
can in principle be obtained by ground state calculations imposing a
non-vanishing crystal momentum. 
By continuity, assuming a normal band insulator,
$\lim_{\kvec \to 0} \min E_{\kvec}(N_e)-E_0(N_e)$ continuously connects
to the optical or vertical gap of the bulk in the zero momentum sector ($\kvec=0$), so that
optical excitation energies at vanishing crystal momentum can be obtained by extrapolation.

Although the conservation of crystal momentum followed from the assumption of a perfect crystalline lattice,
this requirement can be relaxed to include also ionic zero point motion and temperature
effects~\cite{ZPM,HGap}.

\subsection{Twisted boundary conditions}

Twist averaged boundary conditions \cite{Lin01} can be generally used to reduce finite-size errors due
to fermionic shell effects. For a given twist $\theta$, the
many-body wave function is then chosen to obey
\beq
\Psi_{\theta}(\rvec_1,\dots,\rvec_i+L_x \hat{x},\dots)=\re^{\ri \theta_x} \Psi_{\theta}(\rvec_1,\dots,\rvec_i,\dots),
\label{Psitheta}
\eeq
whenever any electron $i$ moves across the simulation cell, e.g., in the $x$-direction.
Electronic eigenstates in the solid can then be labelled by $\theta$ and $\kvec$.

The ground state energy per volume of a solid in the thermodynamic limit 
can then be approximated
by averaging over twisted boundary conditions
\beq
e_0= \frac{1}{M_\theta V} \sum_\theta E_{\kvec_{\theta}} (\theta ,N_{e\theta}),
\eeq
assuming a uniform grid containing
$M_\theta$ twist angles. Here, $E_{\kvec_{\theta}}(\theta,N_{e\theta})$ denotes 
the ground state energy 
of an $N_{e}$ electron system 
at a given twist $\theta$ and crystal momentum $\kvec$, 
 $V$ is the volume of the simulation cell; $N_{e\theta}$
and
$\kvec_\theta$  denote the number of electrons and crystal momentum which minimize
the ground state energy at 
fixed $\theta$.

For a normal band insulator with non-vanishing fundamental gap, the number of electrons 
in the ground state at fixed $N_{e\theta}$ does not
depend on $\theta$; 
canonical and grand canonical twist-averaging give the exact same ground state energy 
in this case \cite{bandgap}.
Although the crystal momentum, $\kvec_\theta$,
 of the ground state wave function 
at a given $\theta$  does not vanish in general,
the total crystal momentum of the ground state
in the thermodynamic limit will be zero, 
$\sum_\theta \kvec_\theta=0$, 
due to time-reversal symmetry.

A non-vanishing total crystal momentum, $\qvec$ can be imposed by 
considering 
a crystal momentum $\kvec_\theta + \qvec$ for the wave function
at a single twist $\theta$. Within canonical twist averaging we can then obtain
the neutral gap by 
\beq
\Delta_n = \min_{\theta,\qvec \ne 0} \left[ 
 E_{\kvec_\theta+\qvec}(\theta,N_e)- E_{\kvec_\theta}(\theta,N_e) \right].
\eeq
However, in contrast to grand canonical twist averaging for the fundamental gap \cite{bandgap},
only a finite set of different crystal momenta, $\qvec$, are accessible in a finite simulation cell.

In a grand-canonical version of the neutral gap, in
addition to changes of the crystal momentum by $\qvec$ at fixed number of electrons
$N_e$,
the number of electrons can vary at different twists with possibly different changes of
the crystal momentum.
For example, we may increase/decrease the 
number of electrons at some $\theta_{1/2}$ compared
to the ground state, $N_e(\theta_{1/2})=N_e \pm 1$.
Now, all values of $\qvec$
become accessible, and the neutral gap will be
the minimum of the canonical one and the fundamental gap, at least to first order,
since energy calculations at different twists are essentially independent.

\subsection{Trial wave functions}

Many-body trial wave functions are conveniently written as
\beq
\Psi_T(\Rvec)=D(\Rvec) \re^{- U(\Rvec)},
\label{PsiT}
\eeq
where $U(\Rvec)$ is totally symmetric under particle exchange, and $D(\Rvec)$ assures
the antisymmetry of fermions, usually in the form of a Slater determinant.
Crystal momentum and twisted boundary conditions, equation~(\ref{Psik}) and equation~(\ref{Psitheta}),
can be imposed by the use of (generalized) Bloch orbitals in the 
Slater determinant
\bea
D(\Rvec) &=& \det_{n \kvec j} \phi_{n \kvec}(\rvec_j | \Rvec),
\\
\phi_{n \kvec}(\rvec |\Rvec) &=& \re^{\ri \kvec \cdot \rvec} u_{n \kvec}(\rvec|\Rvec),
\eea
where $u_{n \kvec}(\rvec |\Rvec)$ may depend symmetrically on all other electron coordinates, $\Rvec$, as it is the case in generalized backflow or neural network orbitals \cite{nbody,BFiter,orbitalBF,NNatoms,Max,MatthewEG}.
If both, $u_{n \kvec}(\rvec |\Rvec)$ as well
as the symmetric factor $U(\Rvec)$, obey periodic boundary conditions
and are invariant
against the translation of all electron positions by crystal lattice vectors,
we have
\beq
\Psi_T(\rvec_1+\tvec,\rvec_2+\tvec, \dots)=
\det_{n \kvec i} \phi_{n \kvec}(\rvec_i +\tvec| \Rvec)
\re^{- U(\Rvec)}
= 
\re^{\ri \sum_\kvec \kvec \cdot \tvec} \Psi_T(\rvec_1,\rvec_2, \ldots).
\eeq
The crystal momentum of the wave function is then given by the sum over 
all occupied Bloch vectors, which must be chosen 
in the grid of wave vectors compatible with $\theta$. Generalizations to include
spin and multideterminant wave functions are straightforward.

\section{QMC results for diamond carbon}\label{sec3}

We have performed electronic QMC calculations on carbon in the diamond structure at ambient pressure $r_s = 1.318$ (lattice constant 3.567 \r{A}). We used a Slater-Jastrow trial wave function, which was fully optimized within variational Monte Carlo, including the long-range (reciprocal lattice) contributions. 
The orbitals in the Slater determinant were taken from DFT calculations with QUANTUM ESPRESSO~\cite{QE2009,QE2017} using the LDA functional with a cutoff of 200 Ry for the kinetic energy and of 400~Ry for the electron
density and with $12\times12\times12$ $\kvec$-point grid. QMC calculations have been performed with the QMCPACK code \cite{QMCPACK}. Burkatzki-Filippi-Dolg pseudopotential \cite{Burkatzki2007} was used to remove the core electrons. Diffusion Monte Carlo calculations were performed with $0.01$ Hartree time step as in~\cite{bandgap}. We used two system sizes: the cubic cell containing eight atoms and a $2\times2\times2$ supercell containing 64 atoms. All calculations are done such that they can be directly compared
with those of the quasiparticle gap of~\cite{bandgap} (see also Supplementary material of \cite{bandgap}).  

\begin{figure}
\center
\includegraphics[width=0.6\columnwidth]{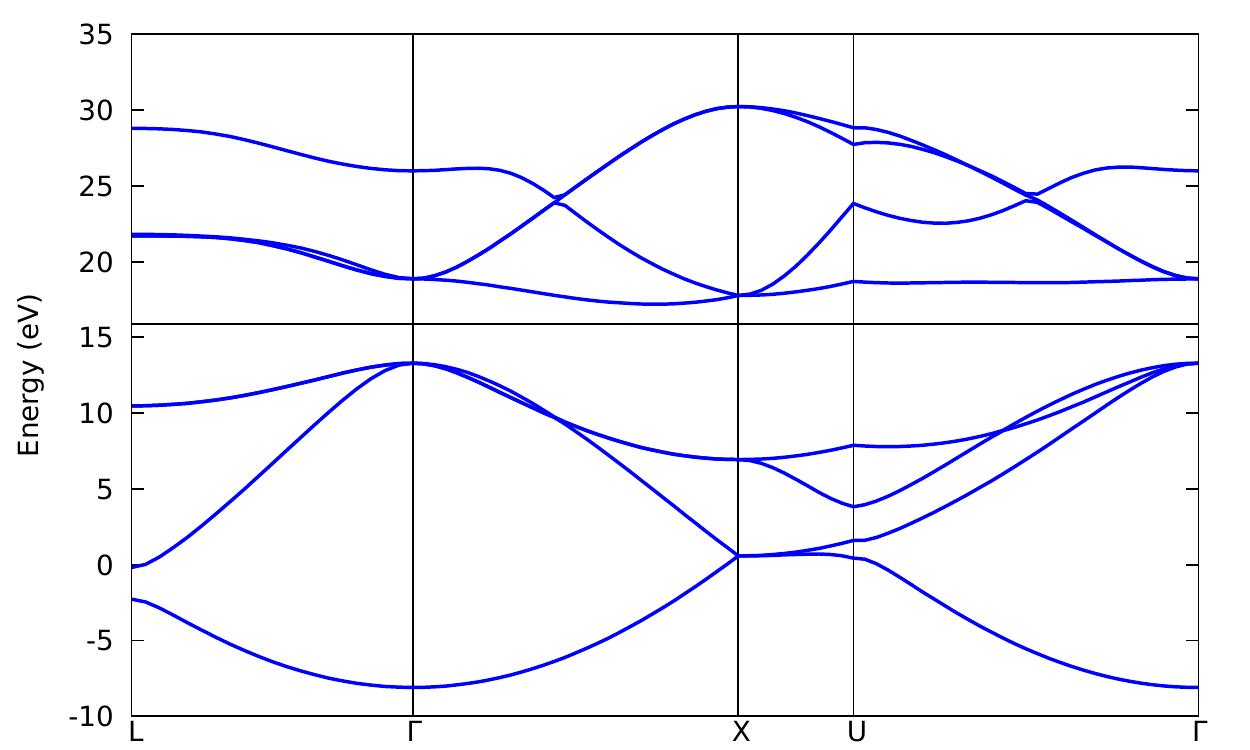}
\includegraphics[width=0.25\columnwidth]{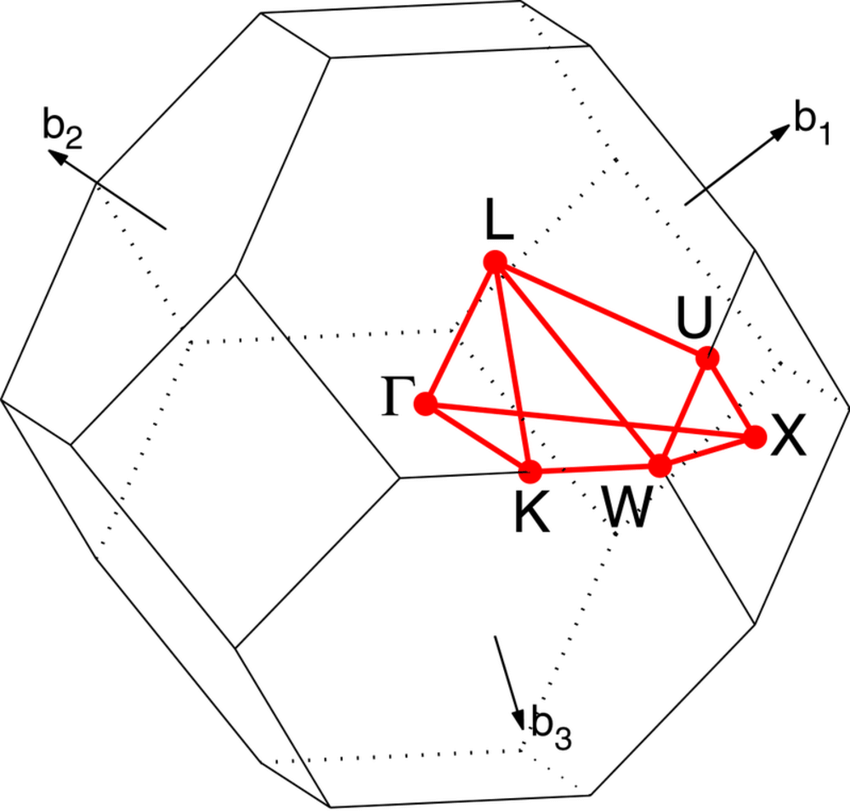}
\caption{(Colour online) Left-hand: DFT-LDA band structure of carbon diamond. Right-hand: Brillouin zone of diamond structure with the selected path for band structure plot.}
\label{fig:diamond_bands}
\end{figure}

Figure \ref{fig:diamond_bands} shows the DFT-LDA band structure of carbon diamond and  the chosen band path in the Brillouin zone. The indirect gap amounts to 3.9 eV. The direct $\Gamma \to \Gamma$ gap is 5.4 eV and the $\Gamma \to X$, which is slightly larger that the indirect gap, is 4.4 eV. We will further limit our QMC calculations to only two neutral electron-hole excitations: $\Gamma \to \Gamma$ and $\Gamma \to X$.
 
\begin{table}[!t]
	\caption{Neutral gap, $\Delta_n$ from 
		DMC calculations of carbon diamond in a supercell containing 
		$N=8$ and $64$ atoms
		for $\Gamma-X$ and $\Gamma-\Gamma$ transitions,
		compared to the corresponding
		quasiparticle band gaps $\Delta_{QP}$ 
		from~\cite{bandgap}. }
    \centering
    \begin{tabular}{c||c |c|c}
    $N$ & $\kvec$  & $\Delta_n$(\kvec) &  $\Delta_{QP}(\kvec)$ \\ 
    \hline
$8$ & $\Gamma \rightarrow X$ & 4.565(6)  & 4.59(2)  \\
& $\Gamma \rightarrow \Gamma$ & 6.265(6)  &  \\
    \hline
$64$ & $\Gamma \rightarrow X$ & 6.04(2)  & 5.98(4)  \\
& $\Gamma \rightarrow \Gamma$ & 7.64(2)  &  \\
    \hline
    \end{tabular}
    \label{tab:energ_gamma_x}
\end{table}

Table~\ref{tab:energ_gamma_x} summarizes our DMC results for the neutral gap
at a fixed wave vector imposed by promoting an electron from $\Gamma$ to $X$, and from $\Gamma$ to $\Gamma$.
We will mainly focus on $\Gamma$ to $X$ in what follows.
The results reported are from single determinant calculations. We have further performed
multi-determinant calculations of the excitations in 
the 8 atom cell, but found no improvement.

For $\Gamma \rightarrow X$, we can compare our neutral gap with the quasiparticle gap
given in the supplementary material of~\cite{bandgap}. We find both gaps to coincide within
our statistical error. Such a coincidence of neutral and quasiparticle gap
for supercells of different sizes has already been
found in diamond silicon \cite{Hunt,Lubos21}.

De facto, these results exclude the use of different
finite-size extrapolation laws for quasiparticle and neutral gap.
Previously, we  derived the asymptotic $1/L$ law for the fundamental quasi-particle
gap as an exact result based on a microscopic analysis with the underlying trial wave function
\cite{bandgap}. Below we will extend this result to the neutral gap and show that the
same $1/L$ asymptotic behavior is recovered as long as the excitations are extended, as it is the case
for our single-determinant wave function.

\begin{figure}
\center
\begin{minipage}[b]{0.49\columnwidth}
\includegraphics[width=\columnwidth]{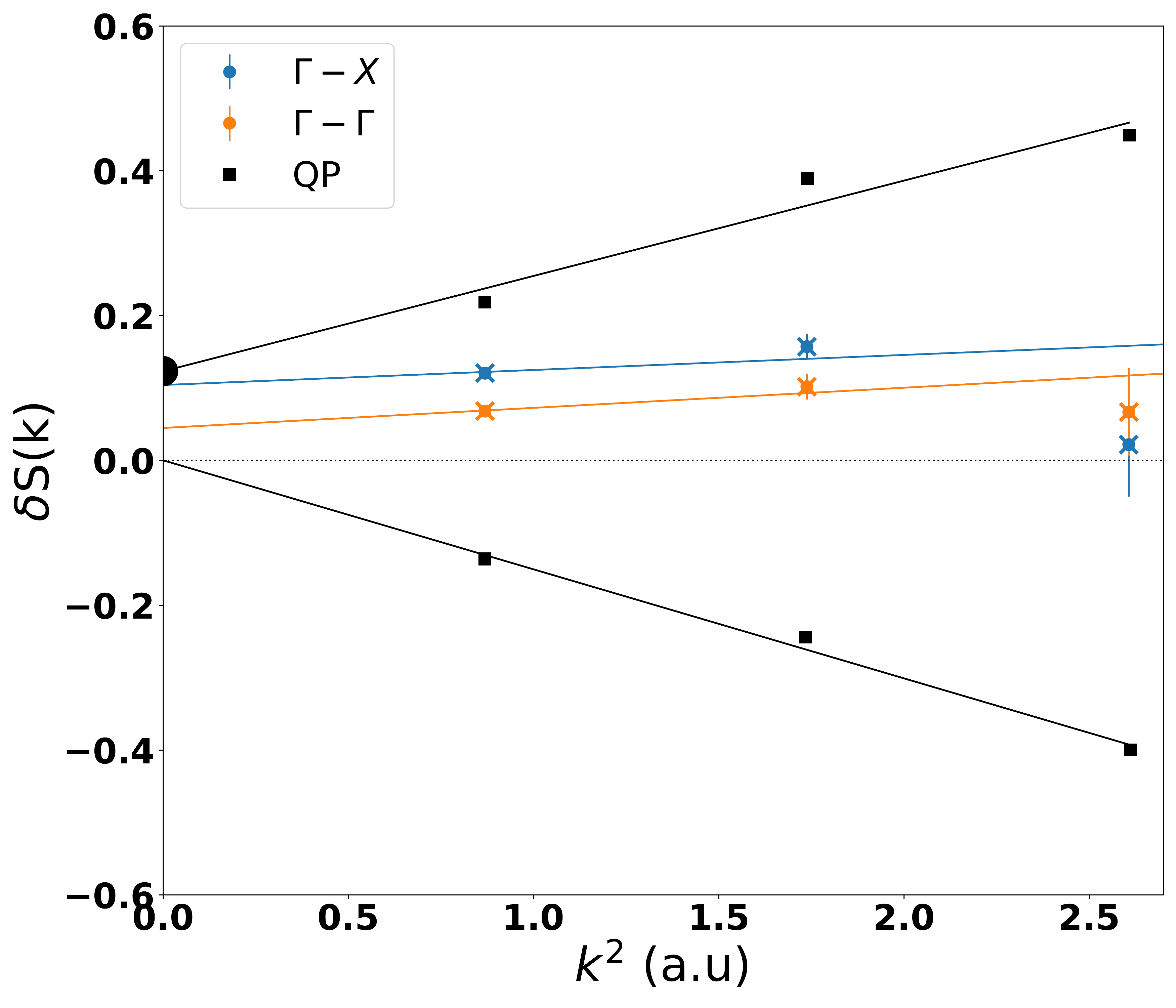}
\end{minipage}
\begin{minipage}[b]{0.49\columnwidth}
\includegraphics[width=\columnwidth]{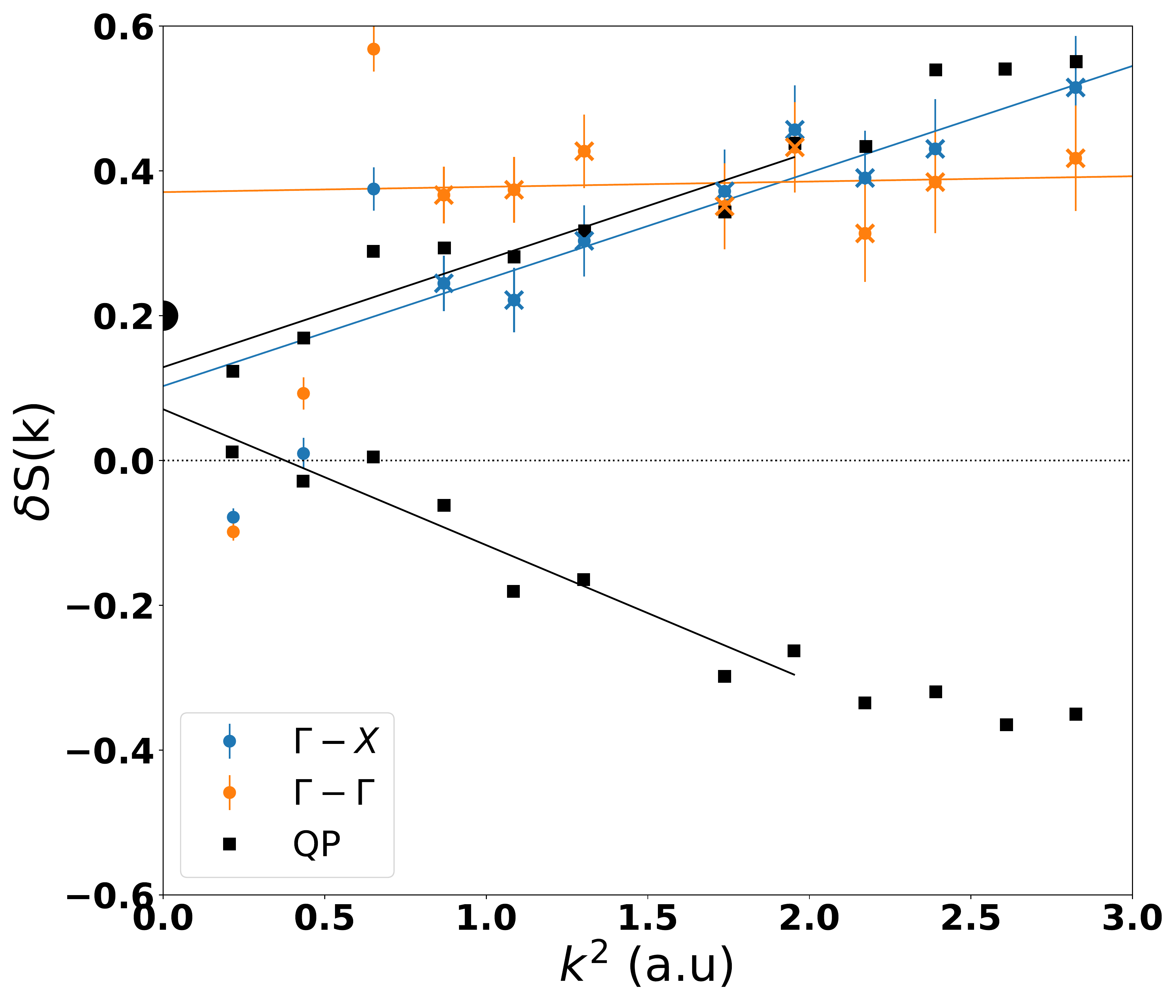}
\end{minipage}
\caption{{(Colour online) Difference between the excited and ground state fluctuating structure factor for carbon, $\delta S(k)$. Blue and orange: neutral excitations for $\Gamma-X$ and $\Gamma-\Gamma$. 
Black: difference of structure factors
from quasiparticle excitations for addition, $\delta S^+(k)$, (upper branch) and removal, $\delta S^-(k)$, (lower branch) from~\cite{bandgap}. Left-hand: 8 atoms, right-hand: 64 atoms. The lines are fits to the data points of the corresponding color. Symbols with crosses are included in the fit. From our anaytical analysis, we expect $\delta S(k) \simeq \delta S^+(k)+ \delta S^-(k)$ for $k \to 0$. Values for small $k$
are likely affected by a larger uncertainty, in particular for
the larger system. Dotted horizontal lines indicate zero. 
}\label{fig:sofk64}}
\end{figure}

\section{Finite size effects for neutral gaps: extended vs localized excitations}\label{sec4}

For the quasiparticle gap, the $1/L$ behavior was directly related to the asymptotic 
$k \to 0$ value\cite{bandgap} of the
structure factor $S(\kvec)= \langle \rho_\kvec \rho_{-\kvec} \rangle/N_e$
where $\rho_\kvec = \sum_j \exp[-\ri \kvec \cdot \rvec_j]$.
In figure~\ref{fig:sofk64}, we compare the difference of the structure factor
of the excited state and the ground state, $\delta S(k)$, from neutral excitation to
the ones obtained from electron addition and removal. Although affected by considerable
noise, and sensitive to the optimisation, especially of the long-range part of the wave function, our results for the neutral gap extrapolate to a non-vanishing value.
The variance of these values at fixed $k$ is large, since different angular directions
seem to follow the envelope given by the electron addition and removal curves (green) with opposite slopes.
The behavior of the structure factor further supports the same $1/L$ extrapolation law
of the quasiparticle gap also for neutral excitations.
Below we analyze the behavior of the structure factor 
for single and multideterminant wave functions and relate
their asymptotic limits to the extended character of the
excitation.

\subsection{Single determinant wave function}

Let us consider a single determinant wave function of the form equation~(\ref{PsiT})
where the Slater determinant is formed out of Bloch orbitals $\phi_{n\kvec}$ labelled
by band index $n$ and crystal momentum $\kvec$. 
In the context of a many-body wave function, two orbitals 
of crystal momentum $\phi_{n\kvec}$ and $\phi_{n\kvec+\qvec}$, $q \to 0$, 
are defined 
to belong to the same band if
they are adiabatically connected in the sense that
\beq
\re^{\ri \qvec \cdot \rvec} \phi_{n \kvec}(\rvec |\Rvec) \simeq \phi_{n \kvec +\qvec}(\rvec | \Rvec),
\label{adia}
\eeq
and $\phi_{n\kvec}(\rvec ) = \phi_{n \kvec+\Gvec}(\rvec)$ is a periodic
function in the reciprocal space of the crystal and we have dropped the dependence
on $\Rvec$ to simplify the notation.
This property is of course verified in the case the orbitals are taken from
any effectively independent electron theory, e.g., Kohn-Sham DFT orbitals as used in our
calculations above.

In a normal band insulator with non-vanishing fundamental gap \cite{bandgap}, we can assume that 
the occupied orbitals (those entering the Slater determinant $D_0$) consist of completely
filled bands, e.g., for each filled band
all of the crystal momentum states within the first Brillouin zone are occupied
and adiabatically connected with each other.  Applying a density fluctuation $\rho_\qvec$
with 
long wavelength, $q \to 0$, we then have
\bea
 \rho_\qvec D_0 &=& \sum_j \re^{\ri \qvec \cdot \rvec_j}
\sum_{n \kvec}^{\text{occ.}}
\frac{\delta D_0}{\delta \phi_{n \kvec}(\rvec_j)} \phi_{n \kvec}(\rvec_j)  
\xrightarrow[q \to 0]{}
\sum_j 
\sum_{n \kvec}^{\text{occ.}}
\frac{\delta D_0}{\delta \phi_{n \kvec}(\rvec_j)} \phi_{n \kvec+\qvec}(\rvec_j)  
\nonumber
\\
&=&
\sum_{j j'} 
\sum_{n \kvec}^{\text{occ.}}
\frac{\delta^2 D_0}{\delta \phi_{n \kvec}(\rvec_j) \delta \phi_{n \kvec+\qvec}(\rvec_{j'})
} \phi_{n \kvec+\qvec}(\rvec_j)  \phi_{n \kvec+\qvec}(\rvec_{j'})
=0,
\eea
where we have used in the second line
that $\phi_{n \kvec+\qvec}$ is already occupied, together with the antisymmetry of the determinant.
Density fluctuations are thus suppressed in the insulating state by the gap in the density of states, and we have $ S_0(q \to 0)=0$ in the ground state.

Let us now consider a particle-hole excitation by promoting an electron from the valence
band $n=v$ at $\kvec_h$ to the conduction band $n=c$ with Bloch momentum $\kvec_p$, described by a Slater determinant
\beq
D_{v\kvec_h \to c \kvec_p}
= \sum_{j}  \frac{\delta D_0}{\delta \phi_{v \kvec_h}(\rvec_j)} \phi_{c \kvec_p}(\rvec_j).
\eeq
Applying again a density fluctuation, we obtain in the long wavelength limit 
\bea
\rho_\qvec D_{v\kvec_h \to c \kvec_p}
&=&
\sum_{j}  \frac{\delta D_0}{\delta \phi_{v \kvec_h}(\rvec_j)} \re^{\ri \qvec \cdot \rvec_j}
 \phi_{c \kvec_p}(\rvec_j)\nonumber\\
&+&  \sum_{i \ne j} 
\sum_{n \kvec \ne v \kvec_h}
\frac{\delta^2 D_0}{\delta \phi_{v \kvec_h}(\rvec_j)  \delta \phi_{n \kvec}(\rvec_i)}
\phi_{c \kvec_p}(\rvec_j) \re^{\ri \qvec \cdot \rvec_i} \phi_{n \kvec}(\rvec_i)
\nonumber
\\
&\xrightarrow[q \to 0]{}& 
\sum_{j}  
\frac{\delta D_0}{\delta \phi_{v \kvec_h}(\rvec_j)}  \phi_{c \kvec_p+\qvec}(\rvec_j)
\nonumber\\
&+& \sum_{i \ne j}
\sum_{n \kvec \ne v \kvec_h} 
\frac{\delta^2 D_0}{\delta \phi_{v \kvec_h}(\rvec_j)  \delta \phi_{n \kvec}(\rvec_i)}
\phi_{c \kvec_p}(\rvec_j) \phi_{n \kvec+\qvec}(\rvec_i)
\nonumber\\
&=& D(v \kvec_h \to c \kvec_p+\qvec) + \frac{1}{D_0} \sum_{i,j} \sum_{n \kvec }
\left[ \frac{\delta D_0}{\delta \phi_{v \kvec_h}(\rvec_j)}
\frac{\delta D_0}{ \delta \phi_{n \kvec}(\rvec_i)}\right.\nonumber\\
&-&\left. \frac{\delta D_0}{\delta \phi_{v \kvec_h}(\rvec_i)} 
\frac{\delta D_0}{ \delta \phi_{n \kvec}(\rvec_j)}
\right] \phi_{c \kvec_p}(\rvec_j) \phi_{n \kvec+\qvec}(\rvec_i) 
\nonumber
\\
&=&D(v \kvec_h \to c \kvec_p+\qvec) - D(v \kvec_h-\qvec \to c \kvec_p).
\eea
We then get for the structure factor
\bea
NS_{v\kvec_h \to c \kvec_p}(\qvec) &=& 
\frac{ \int \rd\Rvec |D_{v\kvec_h \to c \kvec_p}|^2 \re^{-2 U(\Rvec)} \rho_{-\qvec} \rho_{\qvec} }
{\int \rd \Rvec |D_{v\kvec_h \to c \kvec_p}|^2 \re^{-2 U(\Rvec)} }
\nonumber
\\
&\xrightarrow[q \to 0]{}&
\frac{ \int \rd\Rvec |D_{v\kvec_h \to c \kvec_p+\qvec} - D_{v \kvec_h -\qvec \to c \kvec_p}|^2 \re^{-2 U(\Rvec)} }
{\int \rd \Rvec |D_{v\kvec_h \to c \kvec_p}|^2 \re^{-2 U(\Rvec)} }.
\eea
The Slater determinants entering the numerator are orthogonal, so that we have
$NS_{v\kvec_h \to c \kvec_p}(q \to 0) =
N[S^+(q \to 0) + S^-(q \to 0)] =2$
in the noninteracting limit, $U(\Rvec)=0$,
where $S^\pm(k)$ is the structure factor of a quasiparticle excitation with a single electron 
added or removed \cite{bandgap}.
In general, although
 $U(\Rvec)$ will violate orthogonality, the limiting value will not vanish and remains on the order of $S^+(q \to 0) + S^-(q \to 0)$.

 For simplicity, so far, 
 we have assumed that the limiting value $q \to 0$ can be
 taken at fixed number of electrons, neglecting that for finite systems
 the structure factor is only known on a finite grid of $\qvec$.
 Extending the above calculations to the next order, one can see that the
 coefficient of order $q^2$ is extensive $\sim N_e$. 
 However, almost all of the contributions occur equally for ground and excited states, so that the prefactor
 of the $q^2$ behavior cancels to a large extent in the difference,
 $\delta S(q) = N[S_{v\kvec_h \to c \kvec_p}(q)-S_0(q)]$ [and
 similar for $\delta S^\pm(q)$].
Therefore, as in the case of adding or removing electrons \cite{bandgap}, the
difference between the excited and ground state structure factor will remain finite
when approaching $q=0$. 
Assuming that the asymptotic value of $S_{v\kvec_h \to c \kvec_p}$ coincides with $\delta S^+ + \delta S^-$
from the quasiparticle gap, we can
follow exactly the argument given in~\cite{bandgap}.
Then, the leading-order finite-size corrections will be of order $1/L$ and proportional to the
inverse dielectric constant $\epsilon^{-1}$ 
\beq
\Delta_n(\infty)-\Delta_n(L)= \frac{|v_M(L)|}{\epsilon}.
\label{deltaQP}
\eeq

\subsection{Localized excitations: multideterminant wave functions}

Let us now discuss the case of localized excitations, the simplest possibility is via
coherent particle-hole superpositions, e.g.,
\beq
D_e=\sum_\pvec \alpha_\pvec D_{v\kvec_h+\pvec \to c \kvec_p +\pvec}\,,
\eeq
with coefficients $\alpha_\pvec$. We now get
\bea
N S_e( \qvec) 
& \xrightarrow[q \to 0]{} &
\frac{ \int \rd\Rvec |\sum_\pvec
\alpha_\pvec \left[D_{v\kvec_h+\pvec \to c \kvec_p+\pvec+\qvec} 
- D_{v \kvec_h+\pvec -\qvec \to c \kvec_p+\pvec} \right]|^2 \re^{-2 U(\Rvec)} }
{\int \rd \Rvec |\sum_\pvec \alpha_\pvec D_{v\kvec_h+\pvec \to c \kvec_p+\pvec}|^2 \re^{-2 U(\Rvec)} }
\nonumber
\\
&=&
\frac{ \int \rd\Rvec |\sum_\pvec
\left( \alpha_\pvec -\alpha_{\pvec+\qvec} \right) D_{v\kvec_h+\pvec \to c \kvec_p+\pvec+\qvec} 
|^2 \re^{-2 U(\Rvec)} }
{\int \rd \Rvec |\sum_\pvec \alpha_\pvec D_{v\kvec_h+\pvec \to c \kvec_p+\pvec}|^2 \re^{-2 U(\Rvec)} }.
\eea
We now see that the asymptotic value of the structure factor will vanish if the coefficients
$\alpha_\pvec$ become a smooth, differentiable function of $\pvec$ in the thermodynamic limit. 
In this case, $N \delta S(q)$ will vanish as $q^2$,
and the finite-size error of the neutral gap will be of order $1/N$.

\subsection{Excitonic effects}

As already mentioned above, we have not observed any improvement
by the use of multi-determinant wave functions
in the case of diamond in the 8 atom cell.
Still, due to particle-hole attraction, 
localization of the excitation will eventually
occur giving rise to
excitonic effects.

Let us denote by $l_X$ the length scale of localization effects such 
that $|\alpha_p|$ decays exponentially 
with $l_X$ for increasing $p$.
For simulation cells of
extension $L \lesssim 2 l_X$, the decay of $\alpha_p$
is not resolvable due to the finite resolution
$2 \piup/L$ in momentum space. The wave function is then
indistinguishable from a single determinant one, and we have
$S_e(\piup/l_X \lesssim q \lesssim 2 \piup/L) \simeq
S^++S^-$ in this case.
Only for sizes $L \gtrsim 2 l_x$ the
particle hole correlations become effective,
$S_e(q \lesssim \piup/l_x) \sim q^2$.

Assuming a simulation cell
with $L \ll l_X$, 
we see that finite-size effects will still be dominated by the 
approximate flatness of the structure factor until very small wave vectors
$q \lesssim \piup/l_X$ are reached. Until sizes of order of the localization
length are reached, an apparent $1/L$ will remain. 
A simple estimate for finite-size effects including excitonic effects
is given by
\beq
\Delta_n(\infty)-\Delta_n(L)= \frac{|v_M(L)|}{\epsilon} - \frac{|v_M(2 l_X)|}{\epsilon},
\label{deltaGap}
\eeq
where we basically subtract the overshooting
of the correction given in equation~(\ref{deltaQP})
once system sizes of order $2l_x$ are reached
and neglect any further corrections from the
ultimate $q^2$ asymptotics.

Let us stress, that size effects described by equation~(\ref{deltaGap}),
correct for excitonic effects of neutral excitations at the onset
of the continuum. They do not address size effects of calculations
aiming directly at the binding energy of excitons.

\section{Discussion and conclusions}\label{sec5}

We have performed QMC calculations of the neutral gap in diamond carbon
for two supercell sizes.
Our values of the neutral gap was found to practically coincide with
those of quasi-particle gaps of diamond carbon 
for both supercells. Such a quantitative agreement of neutral
and quasiparticle gap has already been observed in
diamond silicon \cite{Hunt}, indicating that finite-size effects are not necessarily different in calculations of neutral or charged gaps.

We have given further analytical arguments for the observed finite-size effects.
Single determinant wave functions imposing exact crystal momentum will necessarily
result in extended, delocalized excitations. The resulting values for the neutral
gap will then obey the same leading order $1/L$ in the thermodynamic limit extrapolation
as in the case of quasiparticle gap calculations.


We have shown how multi-determinant wave functions can describe
localized excitations for a fixed crystal momentum. The localization
ultimately restores a $q^2$ behavior of the structure factor at small wave vectors. This  implies the expected faster $1/N$ convergence 
of the neutral gap,
once the simulation cell exceeds the scale needed to observe localization.

The natural scale of localization is given by the exciton's Bohr radius, $l_X\sim \hbar^2 \epsilon /m_Xe^2$
where $m_X$ is the effective band mass of particle-hole excitations, e.g., the curvature
of the energies involved in the transition. 
This sets also the scale of simulation cells needed to observe
the cross-over from $1/L$ to $1/L^3$ asymptotics in extrapolating neutral gaps.
We have shown how finite-size corrections can be adapted to include
the expected change of the behavior.

In diamond carbon, a rough estimation, based on the measured particle-hole effective mass $m_X\sim0.2m_0$, $m_0$ being a free electron mass, and the dielectric constant $\epsilon=5.7$ \cite{Clark1964}, gives an excitonic length scale of $l_X\sim 28.5$ Bohr, necessitating a simulation cell roughly twice as large.
Using equation~(\ref{deltaGap}) indicating a possible lowering of $\sim |v_M(2 l_X)|/\epsilon =0.24$~eV of the neutral gap 
due to excitonic effects with respect to the
quasi-particle gaps extrapolated with equation~(\ref{deltaQP}).
Such a reduction would also bring the neutral QMC gap closer to experimental values, which however
need to be corrected by electron-phonon coupling effects and the use of pseudopotentials for the
core electrons, see table II of reference~\cite{bandgap}.

Discussing the methodology of QMC calculations for the neutral gap,
we have included
the possibility of twisted boundary conditions. In our QMC calculations on diamond carbon, we have not made use of them, in contrast to previous
calculations of the fundamental gap \cite{bandgap}. 
Changes of the neutral gap with respect to small twists in the
boundary conditions can be used to probe the effective band mass
of particle-hole excitations, such that estimates for $l_X$ can be
obtained without involving knowledge external to the QMC calculations.

\section{Acknowledgements}

The Flatiron Institute is a division of the Simons
Foundation. D. M. C.  is supported by DOE DE-SC0020177. C.P. was supported by the European Union - NextGenerationEU under the Italian Ministry of University and Research (MUR) National Innovation Ecosystem grant ECS00000041 - VITALITY - CUP E13C22001060006.
Computations were done using the Illinois Campus Cluster, supported by the National Science Foundation (Award No. ACI-1238993), the state of Illinois, the University of Illinois at Urbana-Champaign, and its National Center for Supercomputing Applications, and using the GRICAD infrastructure (https://gricad.univ-grenoble-alpes.fr), which is supported by Grenoble research communities. 
This research used HPC resources from GENCI-IDRIS 2022-AD010912502R1 and GENCI (Project No. 544).

\ukrainianpart

\title{Дослідження нейтральної щілини вуглецю методом квантового Монте-Карло}
\author{В. Горєлов\refaddr{label1,label2}, І. Янг\refaddr{label3}, M. Руджері\refaddr{label4}, Д. M. Сеперлі\refaddr{label5}, К. П'єрлеоні\refaddr{label6}, M. Гольцманн\refaddr{label7}}

\addresses{
	\addr{label1} LSI, CNRS, CEA/DRF/IRAMIS, Політехнічна школа, Паризький політехнічний Інститут , F-91120 Палесо, Франція
	\addr{label2} Центр європейської теоретичної спектроскопії (ETSF)
	\addr{label3} Центр обчислюваної квантової фізики, Інститут Флетайрон, Нью-Йорк, Нью-Йорк, 10010, США
	\addr{label4} Центр моделювання, CEA, CNRS, Університет Париж-Сюд, UVSQ, Університет Париж-Сакле, 91191 Гіф-сюр-Іветт, Франція
	\addr{label5} Фізичний факультет, Університет Ілінойса, Урбана-Шампейн, Урбана, Ілінойс 61801, США
	\addr{label6} Факультут фізичних та хімічних наук, Університет Л'Аквіла, Віа Ветойо 10, I-67010 Л'Аквіла, Італія
	\addr{label7} Університет Гренобль Альпи, CNRS, LPMMC, 38000 Гренобль, Франція
}
\newpage
\makeukrtitle

\begin{abstract}
	Представлено метод розрахунку енергетичної щілини зарядово-нейтрального збудження з використанням лише обчислень
	для основного стану. Наведено результати розрахунків методом квантового Монте-Карло $\Gamma\rightarrow\Gamma$ та 
	$\Gamma\rightarrow X$ енергій збуджень ``частинка-дірка'' вуглецю в модифікації алмазу. Проаналізовано вплив скінчених розмірів 
	та виявлено ту ж саму швидкість загасання $1/L$, що і для зарядового збудження, де $L$-лінійний розмір суперкомірки. 
	Таке повільне загасання пояснюється делокалізованим характером збудження в суперкомірках, які надто малі, щоб пристосуватися до ефектів зв’язування екситонів.
	Для систем більшого розміру $1/L$-загасання перетворюється на поведінку типу $1/L^3$. Оцінки довжин екситонного зв'язування
	можна використати для коригування скінченорозмірних ефектів у нейтральних щілинах.	
	\keywords метод квантового Монте-Карло, першопринципні розрахунки, електронна структура, екситони, щілина, алмаз
\end{abstract}

\lastpage
\end{document}